\documentclass[lettersize,journal]{IEEEtran}
\usepackage{amsmath,amsfonts}
\usepackage{algorithm}
\usepackage{array}
\usepackage[caption=false,font=normalsize,labelfont=sf,textfont=sf]{subfig}
\usepackage{textcomp}
\usepackage{stfloats}

\usepackage{verbatim}
\usepackage{graphicx}
\usepackage{cite}
\usepackage{amssymb}
\usepackage{adjustbox}
\usepackage{booktabs}
\usepackage{tabularx}
\usepackage{multicol}
\usepackage[bottom]{footmisc}

\usepackage{algpseudocode}
\usepackage{amsthm,amsmath,amssymb,lipsum}
\usepackage{cleveref}
\usepackage{url}
\usepackage{threeparttable}
\crefname{section}{§}{§§}
\usepackage{listings, xcolor}
\definecolor{verylightgray}{rgb}{.97,.97,.97}

\lstdefinelanguage{Solidity}{
	keywords=[1]{anonymous, assembly, assert, balance, break, call, callcode, case, catch, class, constant, continue, constructor, contract, debugger, default, delegatecall, delete, do, else, emit, event, experimental, export, external, false, finally, for, function, gas, if, implements, import, in, indexed, instanceof, interface, internal, is, length, library, log0, log1, log2, log3, log4, memory, modifier, new, payable, pragma, private, protected, public, pure, push, require, return, returns, revert, selfdestruct, send, solidity, storage, struct, suicide, super, switch, then, this, throw, transfer, true, try, typeof, using, value, view, while, with, addmod, ecrecover, keccak256, mulmod, ripemd160, sha256, sha3}, 
	keywordstyle=[1]\color{blue}\bfseries,
	keywords=[2]{address, bool, byte, bytes, bytes1, bytes2, bytes3, bytes4, bytes5, bytes6, bytes7, bytes8, bytes9, bytes10, bytes11, bytes12, bytes13, bytes14, bytes15, bytes16, bytes17, bytes18, bytes19, bytes20, bytes21, bytes22, bytes23, bytes24, bytes25, bytes26, bytes27, bytes28, bytes29, bytes30, bytes31, bytes32, enum, int, int8, int16, int24, int32, int40, int48, int56, int64, int72, int80, int88, int96, int104, int112, int120, int128, int136, int144, int152, int160, int168, int176, int184, int192, int200, int208, int216, int224, int232, int240, int248, int256, mapping, string, uint, uint8, uint16, uint24, uint32, uint40, uint48, uint56, uint64, uint72, uint80, uint88, uint96, uint104, uint112, uint120, uint128, uint136, uint144, uint152, uint160, uint168, uint176, uint184, uint192, uint200, uint208, uint216, uint224, uint232, uint240, uint248, uint256, var, void, ether, finney, szabo, wei, days, hours, minutes, seconds, weeks, years},	
	keywordstyle=[2]\color{teal}\bfseries,
	keywords=[3]{block, blockhash, coinbase, difficulty, gaslimit, number, timestamp, msg, data, gas, sender, sig, value, now, tx, gasprice, origin},	
	keywordstyle=[3]\color{violet}\bfseries,
	identifierstyle=\color{black},
	sensitive=true,
	comment=[l]{//},
	morecomment=[s]{/*}{*/},
	commentstyle=\color{gray}\ttfamily,
	stringstyle=\color{red}\ttfamily,
	morestring=[b]',
	morestring=[b]"
}

\usepackage{tcolorbox}

\begin{document}

\title{Guardians of the Ledger: Protecting Decentralized Exchanges from State Derailment Defects}

\author{Zongwei Li, Wenkai Li, Xiaoqi Li, Yuqing Zhang
\thanks{Zongwei Li, Wenkai Li, and Xiaoqi Li are with the School of Cyberspace Security, Hainan University, Haikou, 570228, China (e-mail: \{lizw1017, liwenkai871, csxqli\}@gmail.com).} \thanks{Yuqing Zhang is with National Computer Network Intrusion Protection Center, University of Chinese Academy of Sciences, Beijing, 100049, China (e-mail: zhangyq@nipc.org.cn).}
\thanks{Corresponding Author: Xiaoqi Li.}}

\markboth{IEEE Transactions on Reliability}%
{Zongwei Li \MakeLowercase{\textit{et al.}}: Guardians of the Ledger: Protecting Decentralized Exchanges from State Derailment Defects}

\IEEEpubid{}

\maketitle
\begin{abstract}
The decentralized exchange (DEX) leverages smart contracts to trade digital assets for users on the blockchain. Developers usually develop several smart contracts into one project, implementing complex logic functions and multiple transaction operations. However, the interaction among these contracts poses challenges for developers analyzing the state logic. Due to the complex state logic in DEX projects, many critical state derailment defects have emerged in recent years. In this paper, we conduct the first systematic study of state derailment defects in DEX. We define five categories of state derailment defects and provide detailed analyses of them. Furthermore, we propose a novel deep learning-based framework \textsc{StateGuard} for detecting state derailment defects in DEX smart contracts. It leverages a smart contract deconstructor to deconstruct the contract into an Abstract Syntax Tree (AST), from which five categories of dependency features are extracted. Next, it implements a graph optimizer to process the structured data. At last, the optimized data is analyzed by Graph Convolutional Networks (GCNs) to identify potential state derailment defects. We evaluated \textsc{StateGuard} through a dataset of 46 DEX projects containing 5,671 smart contracts, and it achieved 94.25\% F1-score. In addition, in a comparison experiment with state-of-the-art, \textsc{StateGuard} leads the F1-score by 6.29\%. To further verify its practicality, we used \textsc{StateGuard} to audit real-world contracts and successfully authenticated multiple novel CVEs.
\end{abstract}


\section{Introduction}
The DEXs play a crucial role in decentralized finance (DeFi), enabling direct peer-to-peer transactions without intermediaries~\cite{Ramseyer_2023_SPEEDEXa}. Empowered by smart contracts, DEXs facilitate direct interaction between market participants, departing from the traditional reliance on intermediaries like Centralized Exchanges (CEXs)~\cite{XavierFerreira_2023_Crediblea}. In contrast, DEX levarages smart contracts to eliminate the central point of CEX, implementing user-managed assets throughout the transaction process. Thereby, it reduces risks associated with the central exchange being hacked.

With the development of DeFi, DEX suffers from many security issues. DEXs struggle with defects to various attacks, including fund theft, market manipulation, and denial of service~\cite{Xia_2021_Tradea}. For instance, the FixedFloat exchange was exploited by an access control defect with the third-party infrastructure, leading to around \$26 million being stolen in February 2024 \cite{Medium}.

Several previous studies utilized different techniques (e.g., symbolic execution, data invariant detection, chain synchronization)~\cite{chen2019tokenscope, yu2021code, bose2022sailfish, ye2023detecting, invcon2023} to detect state inconsistency vulnerabilities in DEXs. Geoffrey et al.\cite{Ramseyer_2023_SPEEDEXa} introduced SPEEDEX to combat front-running attacks and enhance transaction parallelization in centralized exchanges. In a different approach, Duan et al.\cite{Duan_2022_Automated} developed VetSC, a tool for automated security checks in DApps. Li et al.~\cite{Li_2021_SolSaviour} proposed SolSaviour, a framework for repairing flawed smart contracts. However, despite these advancements, detecting and mitigating state defects in smart contracts remain challenging:

\noindent\textbf{Challenge 1 (C1):} \textbf{Complex State Logic:} The blockchain-based DEX project processes many transactions in a distributed manner, and any related transaction can affect state information. However, due to the complex state logic encapsulated within the DEX contracts, state changes manipulated by an attacker can lead to logic errors in an unpredictable manner. Its higher complexity and interoperability challenge understanding and detecting state changes caused by attacks and malicious behaviors. This is because traditional methods have difficulty capturing complex contracts' structure and interactions when dealing with DEX. They have difficulty in accurately understanding the state changes between contracts.

\begin{figure}[ht]
  \centering
  \includegraphics[width=\linewidth]{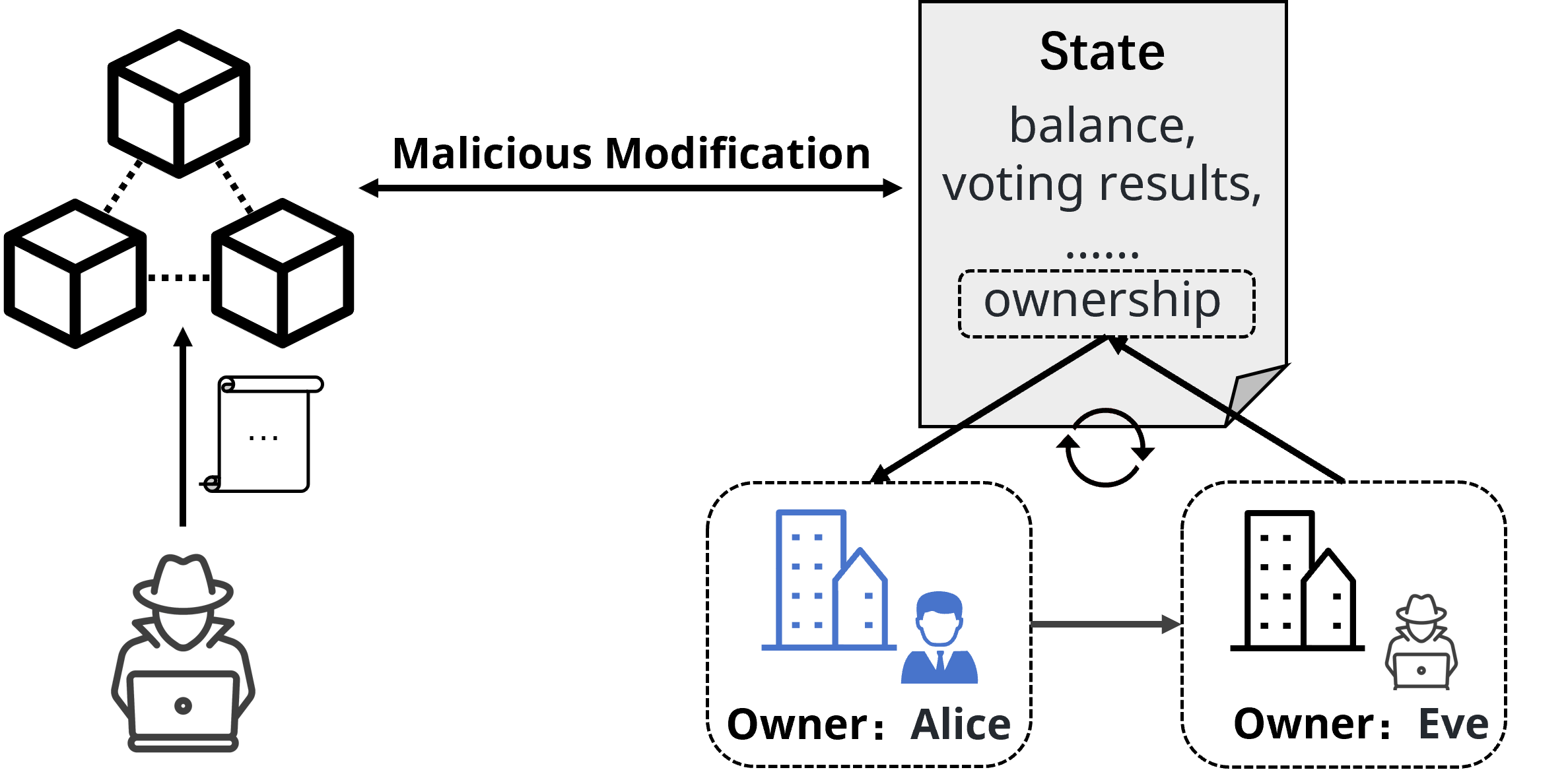}
  \caption{An Illustration of Malicious State Modification.}
  \label{fig:state}
\end{figure}

\noindent\textbf{Challenge 2 (C2):} \textbf{State Derailment Defects:} State derailment defects are distinct from other state defects~\cite{otherdefects} and are a specific category of security defects in DEXs. Fig. \ref{fig:state} represents how these defects can be exploited, resulting in state derailment. These defects originate from various problems, including logical inconsistencies, resource limitations, access control issues, type and declaration errors, and inadequate exception handling. 

State derailment and state inconsistency are critical concepts in understanding the integrity and reliability of system states. State derailment refers to aberrant state behavior where the system deviates from its expected functionality due to unauthorized alterations or erroneous state updates. This deviation can lead to significant disruptions, such as unexpected contract behavior or system failures, thereby compromising the system's reliability and predictability. For instance, a state derailment might occur if a financial transaction system erroneously processes a payment due to an unauthorized state change, causing financial discrepancies. 

On the other hand, state inconsistency pertains to discrepancies or conflicts between various replicas of the same state at the data level. This issue arises when different users or systems access different versions of the same state, leading to a lack of uniformity and potentially causing confusion or errors in decision-making processes. For example, in a distributed database, state inconsistency might occur if one server shows an account balance of \$100 while another shows \$150, leading to conflicting information being presented to users. While state derailment focuses on the deviation from expected behavior due to incorrect state changes, state inconsistency highlights the challenges of maintaining uniformity across multiple state replicas.

\textbf{Our solution:}
To address these challenges, we propose a deep learning-based framework called~\textsc{StateGuard} to analyze complex state logic and detect state derailment defects in DEX smart contracts.

\noindent\textbf{Solution for C1:}
We propose a smart contract deconstructor that can handle various versions of Solidity and AST. It can deconstruct contracts on DEX into ASTs that are easier to analyze. Moreover, it can adaptively extract critical dependency features from the AST, obtaining structural and semantic characteristics of the source code.

\noindent\textbf {Solution for C2: } After extracting the dependency features, we integrate node attributes and critical paths to improve the graph representation. This optimized graph was fed into a GCN model to identify and learn defective features to detect state derailment defects. The GCN not only handles the attribute information of the nodes but also considers the connectivity relationships between the nodes. With the advantages of GCN,~\textsc{StateGuard} can deeply understand the complex logic and state changes in DEX smart contracts, providing more accurate analysis.

The main contributions of this paper are as follows:
\begin{itemize}
\item To the best of our knowledge, we conduct the first systematic study of state derailment defects in DEX contracts. We define five kinds of state derailment defects, which can lead to unauthorized or incorrect modifications to the system state during the execution (\cref{sec:State Derailment Defects}).

\item We propose \textsc{StateGuard}, a novel deep learning-based framework for detecting and analyzing state defects in DEX projects. It learns structural features from the ASTs of DEX contracts and extracts dependent features to identify state derailment defects (\cref{sec:methodology}).

\item We evaluated \textsc{StateGuard} on 46 DEX projects containing 5,671 smart contracts with 94.25\% F1-score. We also conducted a comparative analysis with state-of-the-arts, with the advantages of 6.29\% in F1-score. In addition, \textsc{StateGuard} has discovered multiple novel real-world defects, e.g., CVE-2023-\{47033, 47034, 47035\} (\cref{sec:experiment}).

\item We open source \textsc{StateGuard}'s codes and experimental data at \url{https://figshare.com/s/f44e1399dca60b3672f9}.
\end{itemize}
\section{Background} \label{sec:background}
\subsection{Ethereum and Smart Contract}

The rapid digitalization of society necessitates a secure, efficient, and transparent mechanism for data exchange. With its unique, secure structure of chained data blocks, blockchain technology offers a promising solution. However, early blockchain instances like Bitcoin have limited functionality. Ethereum~\cite{Ethereum}, introduced by Vitalik Buterin in 2013, addresses these limitations by broadening the utility of blockchain through smart contracts. These self-executing programs allow secure, trustless transactions without intermediaries. Ethereum's versatility has led to the development of various Decentralized Applications (DApps), such as those in DeFi, providing users with transparent, secure, and fair services.

A diverse range of DApps, including DeFi applications and DEX, can be developed on the Ethereum platform. These applications leverage the capabilities of smart contracts to furnish users with equitable, transparent, and secure services. However, Ethereum confronts several significant challenges. For instance, due to Ethereum's constrained computational capacity, a high volume of transactions can precipitate network congestion, impacting the user experience. Smart contract security is a concern due to coding defects that may lead to financial losses~\cite{Chen_2020_Survey}. 

\subsection{DApp}
\begin{figure}[htb]
\centering
\includegraphics[width=\linewidth]{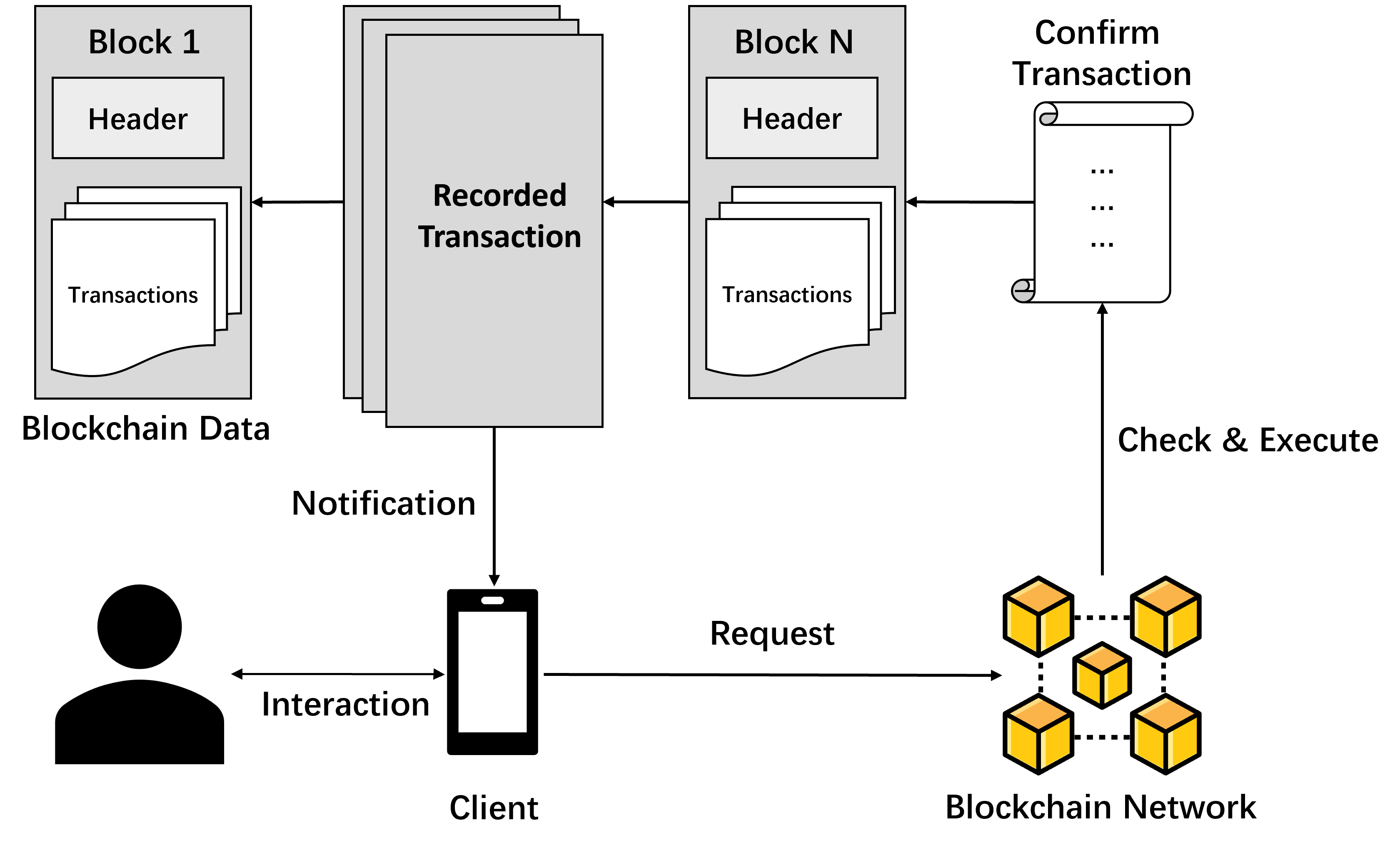}
\caption{Simplified Process of DApp Transaction Requests}
\label{fig:dapp}
\end{figure}

DApp is an application that operates on a blockchain network, free from control by any central authority or individual~\cite{Khan_2020_Pragmatical}. The emergence of this application model offers users an exceptionally secure, transparent, and efficient service platform, enhancing the security and reliability of data storage and transmission. In DApp, users use the application via mobile devices or other clients. They execute smart contracts, record transactions, and verify them on the blockchain network. As depicted in Fig. \ref{fig:dapp}, this process guarantees the decentralization, immutability, and transparency of transactions, embodying the core characteristics of DApp.

The operation of a DApp significantly differs from that of traditional internet applications. Traditional internet applications are governed and maintained by a central server, whereas DApps operate in a decentralized network, with each network node potentially serving as a service provider. This attribute renders the entire system more stable and immune to the damage inflicted by any single node. Leveraging blockchain technology, DApps can ensure data integrity and immutability, significantly bolstering user trust. DApp projects exhibit increased complexity, interoperability, and scalability requirements, thus necessitating greater attention during deployment and management. The development of DApps also confronts challenges, including ensuring application performance, managing a large volume of transactions, and preserving user privacy~\cite{Garamvolgyi_2022_Utilizing}. However, ongoing advancement and refinement in blockchain technology are anticipated to resolve these issues. In summary, DApps constitute a novel application model. Their emergence furnishes us with a more secure, transparent, and efficient service platform, offering boundless possibilities for future application development~\cite{Khan_2020_Pragmatical}.

\section{State Derailment Defects} \label{sec:State Derailment Defects}
In this section, we collect relevant data and define derailment defects.

\subsection{Data Collection}\label{data}

We utilize the DAppSCAN~\cite{zheng2023dappscan}, an open-source dataset of smart contract defects that contains 25,077 smart contracts from 1,139 DApp projects. Considering the financial implications of defects in exchange-type DApps, our study focuses on 46 DApp projects, including 5,671 smart contracts.

\subsubsection{\textbf{Security Analysis Report}}

We are thoroughly analyzing 1,311 security audit reports from 30 specialized entities. Our analysis identifies a category of defects that we refer to as state derailment defects. These defects occur when essential functions fail to maintain the contract's state. Reports frequently underscore how the outcomes of these functions can affect the contract's state but also reveal cases where they either fail to operate as anticipated or are compromised by other functions. We comprehensively investigate smart contracts with these defects, involving a manual review, rigorous analysis, and experimental evaluation. The audit reports are sourced from entities such as ChainSecurity~\cite{chainsecurity}, Runtime Verification~\cite{RuntimeVerification}, Quantstamp~\cite{Quantstamp}, and Smartdec~\cite{Smartdec}.

\subsubsection{\textbf{Data Analysis and Processing}}

We select DEX projects from the DAppSCAN dataset and exhaustively analyze their corresponding audit reports to mark contracts for defects by cross-referencing them. When dealing with DEX-type smart contracts, we find that these contracts are often complex to compile directly. This is mainly because DEX contracts involve complex transaction logic and dependency on other contracts or libraries. Different Solidity versions in one DApp project can cause compilation failures, so it is essential to have strategies for accommodating varying contract versions. In order to compile and process these smart contracts efficiently, we manually process part of the code to ensure that it does not break the original code logic.

\noindent\textbf{Dependency Analysis:} Before compilation, we analyze the dependencies of the smart contracts in the DEX project. It involves identifying the specific versions of contracts and dependency libraries that the contracts depend on. While compiling, we obtain these dependencies and verify the version compatibility to ensure that the source code can be compiled successfully. Specifically, we would need to supplement the missing dependency libraries manually in some special cases.

\noindent\textbf{Version Compatibility:} There is a situation where different smart contracts in the same project could have multiple solidity versions. Therefore, it is necessary to match these smart contracts with the compatible compiler version of solidity.

\noindent\textbf{Syntax and Functional Compatibility:} Due to syntactic and functional differences between versions, there will also be discrepant in dependencies, constructors, etc. Therefore, while modifying dependencies, we would make adjustments to the contracts to ensure compatibility. annotation, during the adjustment process, we do not modify the original semantics and ensure that it is consistent with the defects in the annotation.

\subsection{Defects Definition}\label{definition}
In blockchain and smart contract development, safeguarding contract state integrity is critical~\cite{Dwivedi_2021_Legally}.
State derailment defects constitute a specific category of security defects in smart contracts. Smart contract execution can sometimes result in defects caused by logical inconsistency, design problems, and resource limitations. These defects can cause unauthorized, incorrect, or incomplete updates, changes, or accesses to the system state. Such defects could impact the functionality of a smart contract and may lead to abnormal system operation or exploitation by a malicious user, posing a significant threat to system security~\cite{Sharma_2023_Mixed-Methods}. In smart contracts, the state refers to the stored information or variables representing a contract's current state or status at any given moment. Examples of this data may include account balances, ownership details, or any other relevant information the contract may need to execute its functions properly.

Subsequently, we will dissect five defect categories, examining their roots and potential hazards. Each defect signifies a unique smart contract state issue, highlighting the necessity to comprehend and mitigate these for improved contract security.

\textbf{(1) Logical Inconsistencies}: Logical inconsistencies in the context of smart contracts refer to a mismatch between the actual logic of the contract code and the designer's expectations or intentions, and this inconsistency usually stems from errors, omissions, or defects in the code implementation~\cite{Tchakounte_2022_smart}. It may manifest itself in incorrect updates to the contract state, feature implementations that do not match business requirements, the creation of security vulnerabilities, or incompatibility with other contracts or blockchain platform standards~\cite{Barboni_2023_Smart}. This affects the correctness and security of the contract and can lead to loss of assets or failure of contract functionality.

\begin{figure}[ht]
\setlength{\abovecaptionskip}{-0.cm}
\setlength{\belowcaptionskip}{-0.cm}
\begin{lstlisting}[numbers=none]
function _withdraw(address _account, uint256 _withdrawalID) internal override {
        uint256 amount = withdrawLocks.withdraw(_account, _withdrawalID);
        livepeer.withdrawStake(_withdrawalID);
        steak.transfer(_account, amount);
        emit Withdraw(_account, amount, _withdrawalID);
    }
\end{lstlisting}
\caption{An Example of Logical Inconsistency Defect.}
\label{fig:Logical Inconsistencies}
\end{figure}

\noindent\textbf{Example:}
Fig. \ref{fig:Logical Inconsistencies} displays a smart contract withdrawal function. The function neglects to verify the return value of the \textit{transfer} function, which can cause state derailment if the token transfer operation fails but the contract continues executing subsequent operations. This could compromise the system's functionality, leading to user fund losses or the contract's inability to execute the anticipated logic accurately.

\textbf{(2) Resource Constraints}:
Resource constraints are a series of blockchain resource parameters that constrain the execution of smart contracts, including Gas (execution cost), storage space, network bandwidth, and block time~\cite{Ajienka_2020_empirical}. These constraints have a significant impact on the execution of smart contracts. Developers must consider these constraints when writing smart contracts to optimize code, reduce resource consumption, and improve execution efficiency. If a smart contract cannot be completed due to resource constraints, its state may be impaired, affecting its ability to fulfill its obligations.

\begin{figure}[ht]
\setlength{\abovecaptionskip}{0.cm}
\begin{lstlisting}[numbers=none]
function removePool(address pool_address) public onlyByOwnerOrGovernance {
        require(frax_pools[pool_address] == true, "address doesn't exist already");
        delete frax_pools[pool_address];
        for (uint i = 0; i < frax_pools_array.length; i++){ 
            if (frax_pools_array[i] == pool_address) {
                frax_pools_array[i] = address(0); 
                break;
            }
        }
    }
\end{lstlisting}
\caption{An Example of Resource Constraint Defect.}
\label{fig:Resource Constraints}
\end{figure}

\noindent\textbf{Example:}
In Fig. \ref{fig:Resource Constraints}, the algorithm aims to remove a specified pool address from an associative array cataloging all pool addresses. The code checks for the pool address's existence, removes it from the mapping and sets its array value to 0x0. Notably, the code risks state derailment due to potential gas overconsumption during array traversal for large-scale arrays. 

\textbf{(3)Access Control}:
Access control is a critical mechanism in smart contracts, which defines which users can access or modify the data in the contract, thus protecting the security of the contract~\cite{Ghaleb_2023_AChecker}. By accurately setting access rights, access control ensures that only authorized users or participants can operate on contract data, preventing unauthorized access or modification and maintaining the integrity and security of the contract state. However, there is negligence or error in the design or configuration of the access control mechanism. In that case, it may result in unauthorized users or participants being able to modify the contract data, thus triggering problems with the contract state and affecting the normal execution and security of the contract~\cite{Tolmach_2021_Survey}.

\begin{figure}[ht]
\setlength{\abovecaptionskip}{0.cm}
\begin{lstlisting}[numbers=none]
function claimAirdrop(bytes32 calldata proof[]) {
    bool verified = MerkleProof.verifyCalldata(proof, merkleRoot, keccak256(abi.encode(msg.sender)));
    require(verified, "not verified");
    require(alreadyClaimed[msg.sender], "already claimed");
    _transfer(msg.sender, AIRDROP_AMOUNT);
}
\end{lstlisting}
\caption{An Example of Access Control Defect.}
\label{fig:Access Control}
\end{figure}

\noindent\textbf{Example:}
Fig. \ref{fig:Access Control} exhibits a function handling airdrop claims. It authenticates the claimant's proof and initiates the asset transfer upon successful validation. However, the function neglects to set \verb|alreadyClaimed[msg.sender]| to \textit{true}. This flag prevents users from repeatedly claiming airdrops, but incorrect settings allow for multiple claims, creating security and abuse issues.

\textbf{(4) Type and Declaration Errors}: Type and declaration errors are defects caused by improper variable declarations or incorrect type checking during smart contract development. Such errors are usually detected during the compilation phase, but they can also affect the behavior of the contract at runtime, leading to unexpected state changes. This reduces the security and reliability of smart contracts and exposes them to a high risk of external threats.

\begin{figure}[ht]
\setlength{\abovecaptionskip}{0.cm}
\begin{lstlisting}[numbers=none]
contract TokenVesting {
    uint256 public initialVestAmount; 
    uint256 public vestAmount;
    ...;
    function sendTokens(uint256 _amount) private {
        uint256 vestAmount = _amount;
        if (token.balanceOf(this) < _amount )
        {
            vestAmount = token.balanceOf(this);
        }
        token.transfer(tokenRecepient,vestAmount);
        Vested(vestAmount);
    }
}
\end{lstlisting}
\caption{An Example of Type and Declaration Error Defect.}
\label{fig:Type and Declaration Errors}
\end{figure}

\noindent\textbf{Example:}
Fig. \ref{fig:Type and Declaration Errors} shows a token lock-up contract with a private function called \textit{sendTokens} for sending tokens. Before sending the tokens, the code checks if the balance of tokens in the contract address is sufficient to send the specified amount of tokens. However, there is a potential defect in this code. The function updates a local variable \textit{vestAmount}, but not the contract's public variable of the same name, which could lead to inconsistencies in the contract's state.

\textbf{(5) Exception Handling}: 
In smart contracts, exception handling is a programming mechanism designed to identify and respond to errors or unintended input data during contract execution. By capturing and handling these exceptions, smart contracts can avoid financial losses, contract malfunctions, state update failures, and even security breaches caused by programming errors or improper external inputs, ensuring the stable operation of the contract and the security of the data.

\begin{figure}[ht]
\setlength{\abovecaptionskip}{0.cm}
\begin{lstlisting}[numbers=none]
function approve(address _spender, uint256 _value) 
    public 
    whenNotPaused  
    whenUnlocked 
    returns (bool) 
{
    return super.approve(_spender, _value);
}
\end{lstlisting}
\caption{An Example of Exception Handling Defect.}
\label{fig:Exception Handling}
\end{figure}

\noindent\textbf{Example:}

In the code shown in Fig. \ref{fig:Exception Handling}, if an error occurs while \par \noindent executing \texttt{super.approve(\_spender, \_value)}, it might result in the entire function failing, perhaps leading to the rejection of the entire transaction, which may lead to a denial-of-service attack. Implement robust error handling and fault-tolerance mechanisms to ensure the consistency and integrity of smart contracts.

In essence, state derailment defects exhibit a tendency toward project failures caused by state errors. It underscores vulnerabilities where specific actions lead to the illegal modification of the system's state.

\section{Method}\label{sec:methodology}

In this section, we introduce the principles of \textsc{StateGuard}, which achieves a contract deconstructor and graph optimizer to detect state derailment defects.

\subsection{Overview} 
\begin{figure}[!t]
\centering
\includegraphics[width=\linewidth]{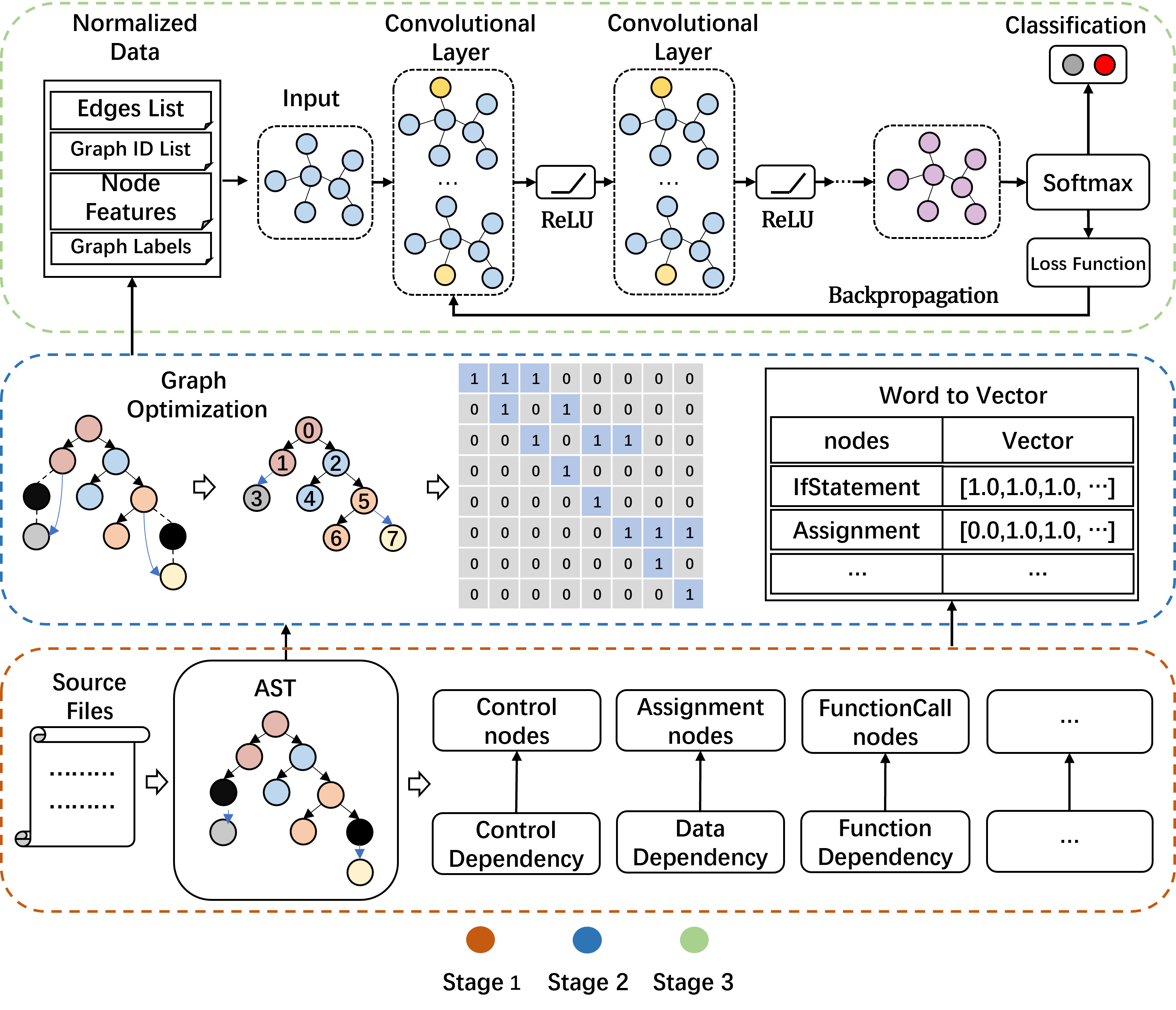}
\caption{Data Processing Workflow of StateGuard.}
\label{fig:module}
\end{figure}

According to Fig. \ref{fig:module}, the overall architecture of our approach consists of three stages:

\noindent\textbf{(1) Smart Contract Deconstructor (\cref{decompiler}):} In this stage, we convert the contracts with different versions into AST in JSON format. It adaptively extracts the contract's dependency to reveal the structural and semantic features of contracts.

\noindent\textbf{(2) Graph Optimizer (\cref{Optimizer}):} In this stage, we integrate node attributes, dynamically identifies and optimizes, and critical dependency paths. Simultaneously, it decomposes contracts into several sub-graphs. Subsequently, it transforms these features from sub-graphs into a standardized data format.

\noindent\textbf{(3) Defect Detection (\cref{GCN}):} In this stage, we feed the standardized data into a GCN for learning potential patterns and features of the graph, and finally identifying the defects.

\subsection{Smart Contract Deconstructor}\label{decompiler}
Smart Contract Deconstructor is designed to process and analyze smart contract code in DEX. It supports different versions of smart contracts and can handle single-contract and multi-contract projects.
The core functionality of the Smart Contract Deconstructor is to convert complex smart contract source code into an easy-to-understand and analyze AST representation and then adaptively identify and extract critical features that significantly impact the contract state.

To improve clarity, we will now break down the process into simpler steps:

\noindent{\textbf{Parsing the Smart Contract:}} The source code of the smart contract is parsed to generate an AST.

\noindent{\textbf{Handling Different Versions:}} The generated AST might have different formats depending on the Solidity version. The deconstructor adapts to these differences to ensure consistency.

\noindent\textbf{Extracting Features:} Critical features that impact the contract state are identified and extracted from the AST.

\begin{figure}[!t]
\centering
\includegraphics[width=\linewidth]{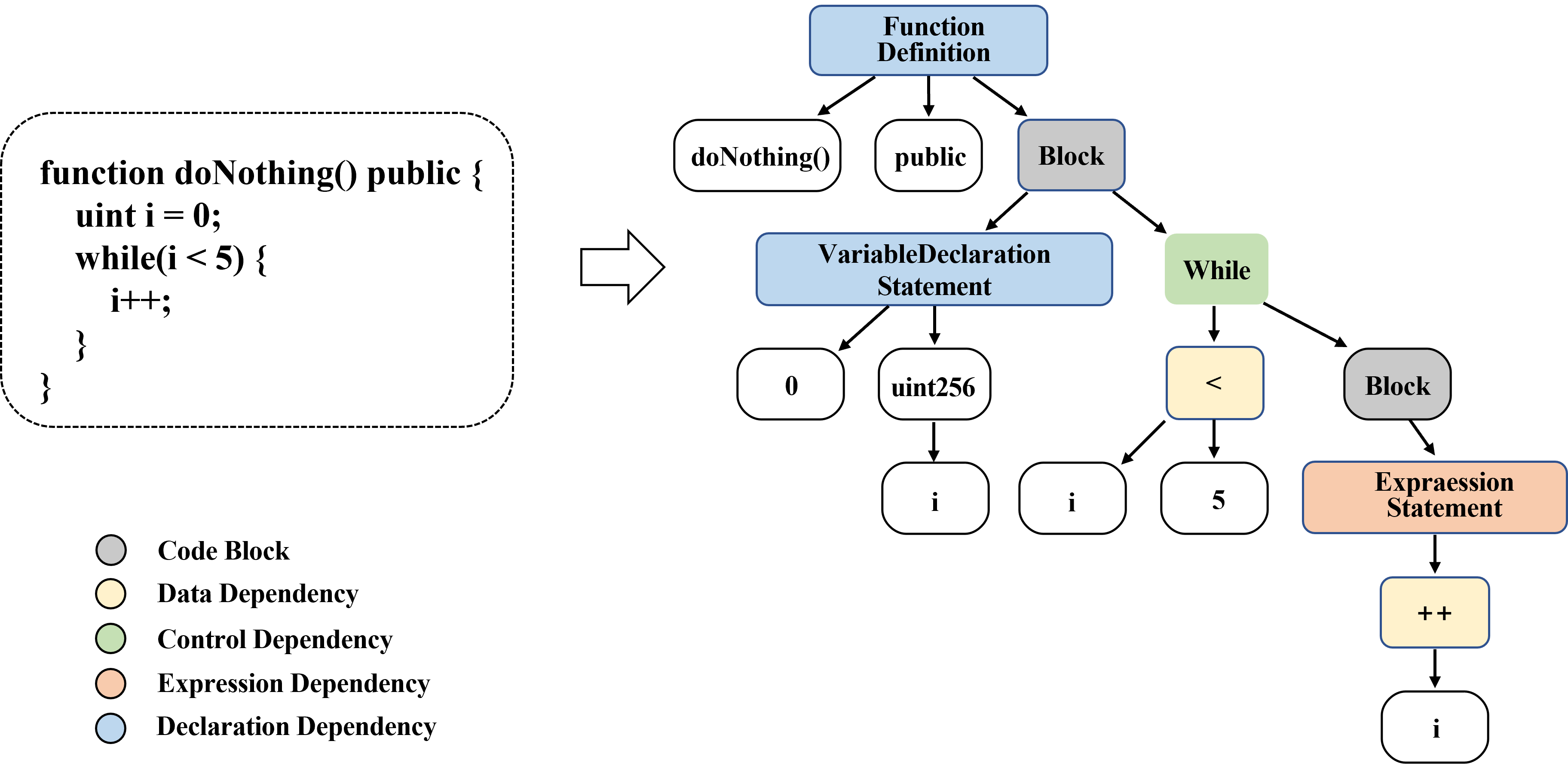}
\caption{An Example of Smart Contract Source Codes' AST.}
\label{fig:ast}
\end{figure}

\subsubsection{\textbf{Abstract Syntax Tree}}
AST is a popular program representation paradigm, effectively encapsulating the semantic relationships between program components~\cite{Wang_2022_Unified}. It is a hierarchical tree structure that represents the source code's architecture, where each node represents a discrete program segment, including functions, declarations, or expressions. 
This representation makes it straightforward to analyze and understand the relationships between different parts of the code, such as function calls, variable declarations, and control flow constructs. By clearly visualizing these relationships, we can more easily pinpoint where issues might arise. 

ASTs provide multiple abstraction layers—ranging from high-level constructs (e.g., functions, loops) to lower-level details (e.g., expressions, operators). This layered abstraction allows for more granular analysis. For instance, higher-level patterns may reveal function-level issues, while lower-level patterns can expose intricate bugs in arithmetic operations or conditional statements.

Research by Curtis J \cite{Curtis2022} has proven that converting programs into ASTs that preserve the semantic relationships of program elements allows for better understanding and manipulation of program logic than bytecode. Inspired by this, we convert smart contracts into ASTs and use this structure to perform in-depth analyses to detect possible security risks and facilitate code optimization.

Fig. \ref{fig:ast} shows a source code fragment and the corresponding AST. The tree's root node represents an entire smart contract or function, each internal node represents a statement or expression, and the leaf nodes represent variables, constants, or other basic elements.
The AST can represent the types of the various nodes in the tree and their corresponding parts in the source code. Examples include function calls and assignment statements. The tree's depth can indicate the code's complexity, while the number of branches in the tree represents the number of decision points in the code. 
ASTs allow us to manipulate the source code at a high level. This manipulability is essential for tasks like feature extraction and graph optimization, making the detection process more efficient.

When we use \texttt{solc} to compile different versions of Solidity smart contracts to generate ASTs, the format of the generated ASTs varies due to the significant differences between Solidity versions. For example, in version 0.6 of the AST, the syntax element of a node is identified in the \textit{name} field, while in version 0.8 of the AST, it is identified in the \textit{nodeType} field. In addition, there is a difference in how child nodes are represented. To overcome this challenge, we analyze the AST format. Despite the formatting differences, each node has a unique \textit{id} attribute, and its children are always in an array. Therefore, we look for an array in the current node to determine if a child node exists. Then, we check whether the \textit{id} attribute is present in this array.

When determining the relationship between nodes, we should follow the following guidelines:
\begin{itemize}
    \item Nodes within the same hierarchy (i.e., nodes in the same array) are considered siblings.
    \item A node in a hierarchy is considered a child of the nearest node in the previous hierarchy (current hierarchy minus one). In other words, if a node is located in a particular level, its parent should be the nearest node in the previous level.
\end{itemize}

\subsubsection{\textbf{Feature Extraction}}
The syntactic features of defect code can be depicted by suitable data structures, which enable us to fetch source code snippets that match these features~\cite{AlDebeyan_2022_Improving}. 
 
We can identify places with security risks by understanding and analyzing how nodes are connected.

Therefore, by traversing the AST, we extract critical features such as control and data dependencies and assign different roles to different nodes to construct the contract graph. Specifically, we defined five types of critical dependency features based on the syntax elements of Solidity.

To clarify, here are the types of critical dependency features:
\noindent{\textbf{Declaration Dependency.}} Variable and constant declaration nodes can represent input, output, and state variables in the code. In contrast, function and method declaration nodes can represent the functionality and operations of the smart contract. Discrepancies here can disrupt state updates, leading to potential state derailment.

\noindent{\textbf{Expression Dependency.}} Syntax and expression nodes encapsulate the logic and computation. The occurrence of defects could interrupt state updates, resulting in state derailment.

\noindent{\textbf{Control Dependency.}} Control dependency nodes define the execution flow of a program, and these operations play an essential role in state management and state changes. Therefore, they are usually highly relevant to defect detection tasks.

\noindent{\textbf{Data Dependency.}} Data dependency refers to the dependence of certain program parts on the state or output of other parts. Understanding and tracking data dependencies is critical in identifying and preventing potential security defects.

\noindent{\textbf{Function Dependency.}} Function dependency refers to the possibility that the behavior of a function may depend on other functions. The main focus is on the relationship between functions and how others influence the behavior of one function.

By extracting these critical dependency features from the AST, we provide the necessary structured information for subsequent dynamic path identification and graph representation simplification.

\subsection{Graph Optimizer}\label{Optimizer}
Most graph neural networks do not consider the important role that specific nodes play in the network during the information transfer process and instead treat all nodes equally. In addition, the complexity of interactions between DEX contracts leads to the generation of overly large and dense graph representations, which increases the computational complexity and thus poses a challenge to the training of graph neural networks. Therefore, to address these issues, we propose a graph optimizer that optimizes the graph representation through, for example, graph attribute integration and dynamic critical path identification to better deal with the performance bottleneck when multiple contracts are encountered.

\subsubsection{\textbf{Graph Attribute Integration}}
An AST is a tree structure representing the source code's abstract syntactic structure. We can construct a directed acyclic graph (DAG) from the extracted node information by traversing the AST. A graphical structure represents the features extracted from the AST, further enabling complex relationships between features.

\noindent\textbf{Node Attribute Integration.} The importance of node building is that it gives a detailed picture of the network structure, reflecting the relationships between individual nodes. This is important for detecting specific defects because understanding and analyzing how nodes are connected can more accurately identify where security risks may exist. For example, if a node has direct connections to numerous other nodes, it could be a prime target for an attack. Attribute information such as  \texttt{id}, \texttt{name}, \texttt{type}, and \texttt{value} of each node is extracted by the smart contract deconstructor. This attribute information is integrated into the node as a node characteristic.

Specifically, we create a set of labels \( L \) to store critical dependencies, which can help us focus on and analyze the necessary nodes for executing defect detection tasks. In addition to node extraction, we preserve node attributes, defined as a tuple \( (N_{\text{id}}, N_n, N_t, N_v) \), denoted as \( w \), where \( N_{\text{id}} \) represents the \textit{unique id} of the node in the syntax tree, \( N_n \) represents the \textit{name} of the node, \( N_t \) represents the \textit{type} of the node, and \( N_v \) represents possible \textit{values} that may exist for the node. This attribute provides a straightforward representation of data, facilitating efficient operations and ensuring immutability.

\noindent\textbf{Edge Attribute Integration.} We further construct edges to model the relationships between nodes. Each directed edge represents a possible traversal path between nodes. Specifically, the features of an edge are extracted as a tuple $(E_s, E_e, E_t)$, where \(E_s\) and \(E_e\) denote its \textit{start} and \textit{end} nodes, respectively, and \(E_t\) denotes the \textit{type} of the edge. Such a construction encapsulates the critical information concisely and clearly, which is easy to manipulate and analyze.

\subsubsection{\textbf{Graph Optimization Strategy}}
After completing the deconstruction of the smart contract, we identify the critical dependency features in the contract and then dynamically analyze the key nodes of the contract and the call relationships between functions. We focus on analyzing those paths with critical dependency features while selectively ignoring those edge paths. By optimizing the dynamically identified critical paths, we simplify the graphical representation by retaining only the critical nodes and connecting edges.

Precisely, our graph optimization process, shown in Fig. \ref{fig:module}, consists mainly of removing nodes and edges that do not belong to a predefined list of specific labels, as well as graph simplification by reducing the graph size. We define a specific list of labels \(L\) and remove all nodes and edges that do not belong to labels in \(L\). In addition, we further simplify and reduce the size of the graph by traversing the graph and removing specific nodes to improve the processing efficiency. The process of graph optimization can be summarised in the following steps:

\begin{itemize}
    \item For each node \(i\) in the graph, if its label \(l_i\) is not in the list of specific labels \(L\), then it is removed, i.e., the optimized set of nodes \(V'=\{i \mid l_i \in L\}\) is obtained.
    \item Performs a depth-first traversal of the optimized set of nodes \(V'\) to establish parent-child relationships between nodes and to remove the ineligible nodes. Specifically, we define \(parent_i\) as the parent of node \(i\) and \(visited\) as the set of visited nodes.
    \item For each node \(i\) in the set \(V'\), if its label \(l_i\) is not in the list of specific labels \(L\) and its parent \(parent_i\) is not null and has been visited (i.e., it is in \(visited\)). If the node \(i\) has no child node \(j\), then the node \(i\) and its connection edge \((parent_i, i)\) with its parent \(parent_i\) are removed. If node \(i\) has child \(j\), remove node \(i\) and its connecting edges \((parent_i, i)\) and \((i, j)\) with its parent \(parent_i\) and child \(j\), and set the parent of \(j\) to \(parent_i\) and create a new connecting edge \((parent_i, j)\) with its key children inherit to the parent node.
\end{itemize}

Through the above steps, we use the optimized set of nodes and edges as the set of nodes and edges of the final optimized graph.
Furthermore, for multi-contract projects in DEX, we first identify the nodes that interact between different contracts (i.e., function dependency) and represent these contracts as subgraphs. We then merge these subgraphs into a complete graph. Doing so enables the graph to represent critical dependencies more centrally, thus improving the accuracy and efficiency of defect detection.

\begin{algorithm}[t]
\caption{Source Code to Normalized Data}
\label{alg:alg1}
\begin{algorithmic}[1]
\Procedure{SourceToGraph}{$source\_file$}
    \State $AST \gets \text{Parse}(source\_file)$
    \State $word2idx, M \gets \text{Preprocess}(AST)$
    \State $A, N, V \gets \text{ASTtoAdjMatrixAndDict}(AST, word2idx)$
    \State $G \gets \text{OptimizeGraph}(A, N, V)$
    \State $G' \gets \text{Normalize}(G)$
    \State \Return $G'$
\EndProcedure

\Procedure{\small ASTtoAdjMatrixAndDict}{$AST, word2idx$}
    \State $A, N, V \gets \text{Initialize empty matrix and dictionaries}$
    \For{each $node$ in $AST$}
        \State $w_i \gets \text{node.attributes}$
        \State $x_i \gets M[:, word2idx(w_i)]$
        \State $V[i] \gets x_i$
        \For{each $neighbor$ in $node.neighbors$}
            \State $A[i, word2idx(neighbor.label)] \gets 1$
            \State $N[i] \gets neighbor$
        \EndFor
    \EndFor
    \State \Return $A, N, V$
\EndProcedure

\Procedure{OptimizeGraph}{$A, N, V$}
    \State $V' \gets \text{RemoveNodesAndEdges}(A, N, V)$
    \State $G \gets \text{DFSAndRemove}(V')$
    \State \Return $G$
\EndProcedure

\Procedure{Normalize}{$G$}
    \State $G' \gets \text{ApplyTransformations}(G)$
    \State \Return $G'$
\EndProcedure
\end{algorithmic}
\end{algorithm}

\subsubsection{\textbf{Graph Embedding}}
Word2Vec~\cite{Church_2017_Word2Vec} is a neural network language model that can transform text data into vector format by learning semantic relationships between words. These vectors can provide valuable inputs for various deep-learning tasks, including text classification, sentiment analysis, and information retrieval. We convert each graph into adjacency matrices and dictionaries, with node labels as feature vectors. We map node labels into feature vectors by employing the \(word2idx\)~\cite{Asudani_2023_Impact} dictionary and an embedding matrix \(M\). The vector representation of each node \(i\) is \(v_i = E_{word2idx}(l_i)\). The topology of a graph \(G=(V,E)\) is depicted using an adjacency matrix \(A\). For each node \(i\), an adjacency dictionary \(N_i\) is constructed to denote its directly adjacent nodes. In node representation learning, we aim to derive each node's representation vector \(h_{i}\), mapping each node's feature vector to the representation space via a non-linear transformation, \(h_i = f(X_i)\). This process is crucial to processing graph data, enhancing its applicability and efficiency.

\subsubsection{\textbf{Graph Normalization}}
The trimmed node and edge sets form the refined graph, which is then normalized. Each node attribute \(w_i\) is associated with a feature vector \(\mathbf{x}_i\) and further processed to normalized vectors \(\mathbf{z}_i\). Algorithm \ref{alg:alg1} outlines the process of source code to normalised data.

The normalized data, including node lists, edge lists, node features, and graph labels, are readied for ensuing deep-learning tasks. Inter-node relationships are preserved in an edge list \(E'=\{(i,j)|A_{ij}=1, i, j \in V'\}\). Each node's id is stored in a list \(G'=\{g_i | i \in V'\}\). Node labels, features, and graph labels are preserved in corresponding lists. The adjacency matrix and dictionary encapsulate the graph's connections. Significantly, we have developed an automated tool to convert source code into normalized data, so the entire process is fully automated.

\subsection{Defect Detection Based on GCN}\label{GCN}
When dealing with graphical data, the normalized data can act as the input for GCN. GCN~\cite{Yu_2021_Knowledge} is a crucial algorithm for processing graph-structured data, learning, and generating vector representations of nodes by iteratively propagating node features. 
The key design principles behind our GCN-based approach are as follows:

\noindent{\textbf{Graph Representation of Smart Contracts:}} We represent smart contracts as graphs, where nodes represent syntactic constructs and edges represent syntactic relationships between these constructs. This allows us to leverage GCN's ability to process graph-structured data effectively.

\noindent{\textbf{Local Connectivity and Feature Aggregation:}} GCN excel at capturing local neighborhood information in graphs. By iteratively aggregating information from neighboring nodes, GCN can build rich representations of node contexts, capturing semantics derived from both local and global structures.

\begin{figure}[!t]
\centering
\includegraphics[width=\linewidth]{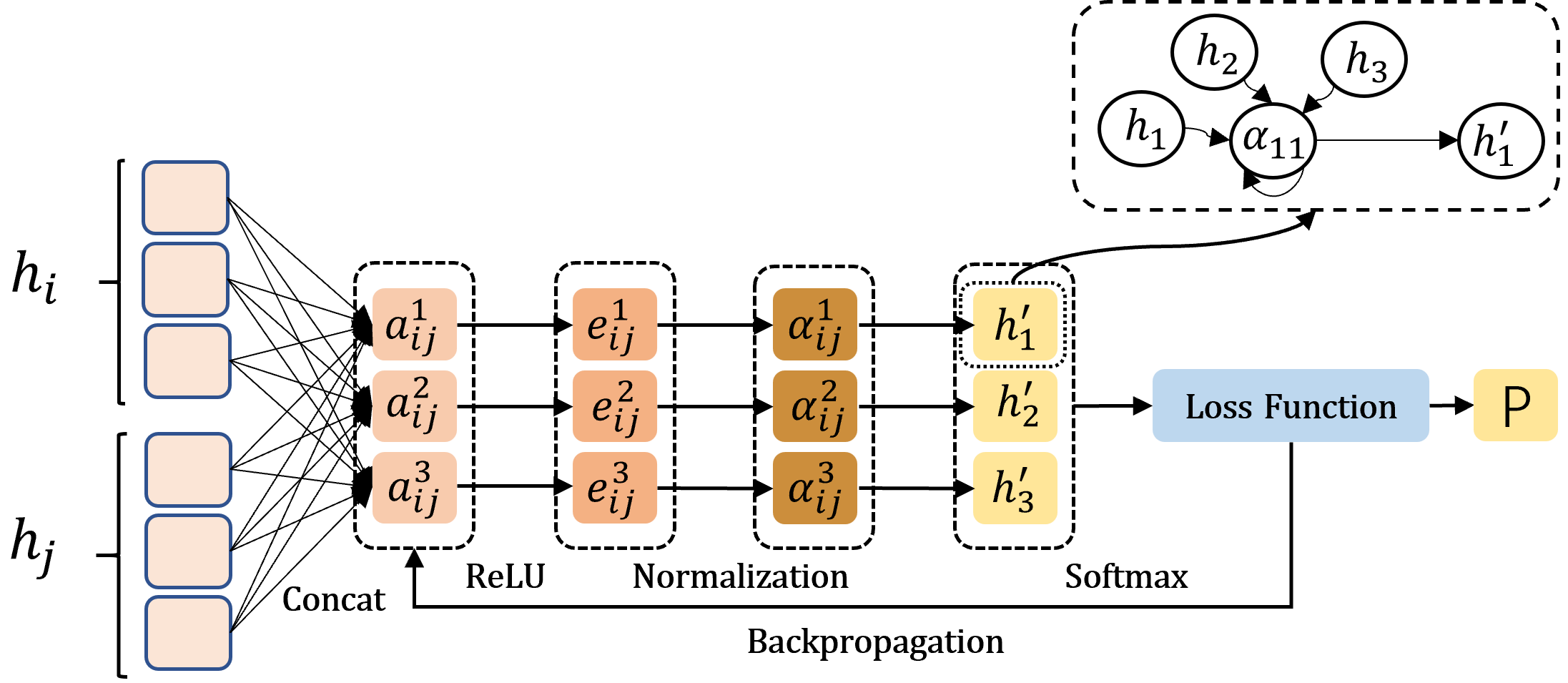}
\caption{Graph Convolutional Neural Network.}
\label{fig:gcn}
\end{figure}

\noindent{\textbf{Hierarchical Information Propagation:}} By stacking multiple convolutional layers, GCN can capture hierarchical information effectively. Each layer captures higher-level semantics by combining information from lower-level layers, providing a comprehensive understanding of the smart contract code. As shown in Fig.~\ref{fig:gcn}, the GCN-based approach effectively captures hierarchical information through multiple convolutional layers.
GCN learns node representations by propagating node features in the form of:
\begin{equation}
\begin{aligned}
    H^{(l+1)} = \sigma(\hat{D}^{-\frac{1}{2}}\hat{A}\hat{D}^{-\frac{1}{2}}H^{(l)}W^{(l)})
\label{eqa:gcn}
\end{aligned}
\end{equation}

Each layer utilizes the normalized adjacency matrix with self-loops $\hat{A}$ and the diagonal matrix of degree plus self-loops $\hat{D}$ to propagate node features. \cref{eqa:gcn} updates the node feature matrix $H^{(l+1)}$ by multiplying the weight matrix $W^{(l)}$ and applying the activation function $\sigma$. It can be interpreted as weighting the sum of node features and the features of its neighboring nodes and then performing a non-linear transformation through the activation function.

The output prediction employs the softmax function \cref{eqa:softamx} to make predictions by multiplying the node feature $H^{(L)}$ of the last layer with the weight matrix $W^{(L)}$. We can calculate it as following \cref{eqa:softamx}:

\begin{equation}
\begin{aligned}
    \hat{y}_i = \text{softmax}(H^{(L)}W^{(L)})
\label{eqa:softamx}
\end{aligned}
\end{equation}

\noindent where $L$ indicates the last layer, and $\hat{y}_i$ denotes the predicted result.

In order to improve the model's performance, we use the backpropagation algorithm~\cite{Ganin_2015_Unsupervised} along with an optimizer to update the model's parameters. For binary classification problems, the cross-entropy loss function can be used to gauge the difference between the predicted output and the actual label. The model is optimized by minimizing this loss function as following \cref{eqa:loss}: 
\begin{equation}
\begin{aligned}
   \text{Loss} = -\frac{1}{N}\sum_{i=1}^N\left[y_i\log(\hat{y}_i) + (1-y_i)\log(1-\hat{y}_i)\right]
\label{eqa:loss}
\end{aligned}
\end{equation} 

\noindent where $y_i$ represents the actual label of the sample $i$, and $\hat{y}_i$ is the model's prediction for the sample $i$.

\section{EXPERIMENT} \label{sec:experiment}
In this section, we conduct a series of experiments, to evaluate the effectiveness of \textsc{StateGuard}.

\subsection{Experimental Settings}
All experiments are executed on a Ubuntu server 22.04 LTS equipped with NVIDIA GeForce GTX 4070Ti GPU, Intel(R) Core(TM) i9-13900KF CPU, and 128G RAM. The software environment includes Python 3.9 and PyTorch 2.0.1.

Regarding the model configuration, we utilize a three-layer GCN, set the adaptive learning rate, and choose the \texttt{ReLU} function as the activation function. During model training, we use cross-entropy as the loss function and \texttt{Adam} as the optimization algorithm. We use 90\% of the dataset for training and the remaining 10\% for validation. In order to comprehensively evaluate the performance of the model on the test set, we select accuracy (ACC), recall, precision, F1 score, and false positive rate (FPR) as the evaluation metrics. 

Specifically, \textbf{ACC} is the ratio of correct predictions to total instances. \textbf{Recall} measures how many actual positives we capture. \textbf{Precision} reflects how many positives are truly positive. \textbf{F1 Score} balances Precision and Recall. \textbf{FPR} indicates how often negatives are incorrectly identified as positive.

\subsubsection{\textbf{Dataset}}\label{Dataset}
We use the publicly available DAppSCAN dataset \cite{zheng2023dappscan} to build a comprehensive dataset critical for identifying and analyzing defects in DApp projects. The dataset includes 703 DApp projects, totalling 23,637 smart contracts, and is continuously updated. We selected 46 DEX projects for analysis, which include a total of 5,671 smart contracts. In addition, we collected 1,311 security analysis reports from 30 companies or organizations conducting security audits of blockchain technology, smart contracts, and cryptocurrency projects. We thoroughly analyzed the defects in these reports, cross-referencing them with smart contracts for our experimental purposes. We also used another publicly available dataset, Smartbugs~\cite{DurieuxEtAl2020ICSE}, a traditional smart contract dataset containing 4,285 smart contracts. Table \ref{table:dataset} shows the number of smart contracts we used, containing both vulnerable and benign contracts.

\begin{table}[h!]
\centering
\caption{The Collected Dataset for Our Evaluation. \# indicates the number of each item.}
\begin{tabular}{>{\centering\arraybackslash}p{2.2cm} c c}
\toprule
\textbf{Dataset} & \textbf{\# Contracts} & \textbf{ \# Audit Reports} \\
\midrule
DAppSCAN & 5,671 & 1,311 \\
\midrule
Smartbugs & 2,000 & 0 \\
\bottomrule
\end{tabular}
\label{table:dataset}
\end{table}

\textbf{Evaluation Metrics.} The effectiveness of \textsc{StateGuard} is evaluated based on the following research questions (RQs):

\begin{itemize} 
\item \textbf{RQ1:} Is \textsc{StateGuard} capable of accurately identifying state derailment defects in the public dataset?
\item \textbf{RQ2:} Can \textsc{StateGuard} find state-related defects undetectable by other tools? How does it compare with existing tools?
\item \textbf{RQ3:} Can \textsc{StateGuard} effectively detect defects in real-world contracts?
\end{itemize}

\subsection{Answer to RQ1: Defects Detection in a Large-Scale Dataset} 
To address RQ1, we conduct experiments on 5,671 smart contracts from DAppSCAN. We use 90\% of these for training and the remaining 10\% for testing. The experimental results presented in Table \ref{tab:table1} depict the performance of \textsc{StateGuard}, including ACC, Recall, Precision, F1-score, and FPR. \textsc{StateGuard} only identifies whether the contract contains a defect, so we only count it once even if the defect occurs multiple times. As illustrated in Table \ref{tab:table1}, these results substantiate the superior performance of \textsc{StateGuard} in detecting state derailment defects.

\begin{table}[htbp] 
\centering 
\caption{Performance Metrics of StateGuard.} 
\label{tab:table1} 
\resizebox{\columnwidth}{!}{
\begin{threeparttable}
\begin{tabular}{@{}lccccc@{}}
\toprule 
\textbf{Tool} & \textbf{ACC(\%)} & \textbf{Recall(\%)} & \textbf{Precision(\%)} & \textbf{F1(\%)} & \textbf{FPR(\%)} \\
\midrule 
StateGuard & 94.83 & 94.82 & 98.28 & 94.25 & 0.03 \\
\bottomrule 
\end{tabular} 
\end{threeparttable}
}
\end{table}

The experimental outcomes demonstrate the efficiency of \textsc{StateGuard} in detecting state derailment defects. The detection ACC of \textsc{StateGuard} reaches 94.83\%, and the recall rate is 94.82\%, indicating that it can accurately identify the most defects. The Precision reaches up to 98.28\%, showing that most defects identified by \textsc{StateGuard} are indeed actual. Moreover, the F1-Score is 94.25\%, reflecting the comprehensive performance of \textsc{StateGuard}. The FPR is only 0.03\%, showing that \textsc{StateGuard} rarely mislabel standard cases containing defects. It shows that \textsc{StateGuard} demonstrates performance in detecting state derailment defects, characterized by high accuracy, high recall, and markedly low FPR.

\begin{tcolorbox}[boxrule=1pt,boxsep=1pt,left=2pt,right=2pt,top=2pt,bottom=2pt]
\textbf{Answer to RQ1.}
The results indicate that \textsc{StateGuard} can identify state derailment defects in the public dataset with considerable accuracy and low false positive rates.
\end{tcolorbox}

\subsection{Answer to RQ2: Comparison Experiment} 
In response to RQ2, most detection tools only analyze individual contracts and cannot perform comprehensive detection on the entire project, while multi-contracts in DApps are often more complex. This complexity arises from the interaction and dependency between multiple contracts, for which existing tools provide limited support. Furthermore, although some tools can support dependency imports, due to differences between these tools, they require separate adaptation and adjustment for each tool, which is not only time-consuming but also prone to errors.
To ensure the validity of the experimental outcomes, we follow the action in~\cite{Yang_2023_Definition}. We employ a random sampling strategy to select 2,000 smart contracts exhibiting state derailment from the SmartBugs dataset. Concurrently, we gather a series of smart contract defect detection tools from renowned journals and conferences in the fields of software and security (e.g., CCS and ASE), as well as Mythril~\cite{mythril}, which is recommended by the official Ethereum community.

To facilitate comparative analysis, we select eight benchmark smart contract detection tools, i.e., Mythril, Oyente~\cite{Luu_2016_Making}, Securify~\cite{tsankov2018securify}, Confuzzius~\cite{Torres_2021_ConFuzzius}, Conkas~\cite{Veloso__Conkas}, Manticore~\cite{Mossberg_2019_Manticore}, Slither~\cite{Feist_2019_Slither}, and Smartcheck~\cite{Tikhomirov_2018_SmartCheck}. During the selection process, we consider several factors: \underline{(1)} the accessibility of the tool's source code; \underline{(2)} the tool's capability to detect defects related to the contract state; \underline{(3)} the tool's support for source code written in Solidity; \underline{(4)} the tool's ability to report the specific locations of potential defect code for manual review.

The experimental results are presented in Table \ref{tab:performance_comparison}. As with RQ1, \textsc{StateGuard} only identifies whether the contract contains a defect, so we only count it once if the defect occurs multiple times. Smartcheck indicates that almost all contracts have defects in the dataset. Meanwhile, tools such as Slither and Manticore fail to provide results due to compilation errors, timeouts, or their inability to handle some of the latest versions of smart contracts. In addition, apart from Securify and \textsc{StateGuard}, the other tools fail to process all contracts. The analysis results of each tool are filtered, retaining only the valid results to ensure the fairness of the analysis.

\begin{table}[!t]
\centering
\caption{Performance Comparison of Related Tools.} 
\label{tab:performance_comparison} 
\resizebox{\columnwidth}{!}{
\begin{tabular}{@{}lccccc@{}} 
\toprule
\textbf{Tools} & \textbf{ACC(\%)} & \textbf{Recall(\%)} & \textbf{Precision(\%)} & \textbf{F1(\%)} & \textbf{FPR(\%)} \\ 
\midrule
Mythril     & 34.89   & 47.34      & 50.26         & 48.75        & 88.67   \\
Confuzzius  & 53.43   & 53.44      & 66.03         & 59.07        & 46.59   \\
Oyente      & 52.53   & 50.15      & 90.67         & 64.58        & 32.48   \\
Securify    & 74.10    & 56.90       & 86.74         & 68.72        & 8.70     \\
Conkas      & 74.72   & 81.10       & 89.34         & 85.02        & 74.13   \\
\textbf{StateGuard} & \textbf{91.40}    & \textbf{90.40}       & \textbf{92.24}         & \textbf{91.31}       & \textbf{7.60}     \\
\bottomrule
\end{tabular}
}
\end{table}

In a comparison experiment, \textsc{StateGuard} demonstrates performance metrics. Table \ref{tab:performance_comparison} shows that it outperforms other detection tools in several essential performance metrics. Notably, \textsc{StateGuard} has achieved an accuracy rate of 91.40\% and a recall rate of 90.40\%. At the same time, its Precision is 92.24\%, and the F1 score reaches 91.31\%, both showing excellent performance. It is particularly noteworthy that the false positive rate is only 7.60\%, significantly reducing the probability of false positives.

The experimental findings indicate that \textsc{StateGuard} can discover state-related defects that other tools may miss and exhibit a significant advantage in its overall performance.

\begin{tcolorbox}[boxrule=1pt,boxsep=1pt,left=2pt,right=2pt,top=2pt,bottom=2pt]
\textbf{Answer to RQ2.}
The experimental results demonstrate that \textsc{StateGuard} can identify unique state-related defects that other tools may overlook, and it can also surpass these tools in several metrics (i.e., accuracy, precision, recall), confirming its efficacy in complex multi-contract environments.
\end{tcolorbox}

\subsection{Answer to RQ3: Real-world Contract Detection} 
We randomly select 1,596 samples of smart contracts from Etherscan~\cite{etherscan}, which cover smart contracts of different sizes. The sampling methodology we adopt ensures the applicability and validity of our research results.

We run \textsc{StateGuard} to detect these real-world smart contracts, and the results show that \textsc{StateGuard} successfully identifies smart contracts with state derailment defects. We apply for and obtain Common Vulnerabilities and Exposures (CVE) certifications for CVE-2023-47033, CVE-2023-47034, and CVE-2023-47035. This means that they are recognized security defects that malicious users could exploit. These defects have been publicized and notified to the vendor. We have also submitted a detailed security audit report to Etherscan that includes the smart contract address with the defect, the exact location of the defect, its property, and the potential impact. It is worth noting that these defects are not successfully detected when using the benchmark tool in RQ2. 
This indicates that \textsc{StateGuard} is more effective in detecting state derailment defects.
The advantage of \textsc{StateGuard} is its ability to capture the interaction of functions and state variables in smart contracts through graph structures. In addition, GCN can learn the structural features of graphs that are difficult to capture by traditional static analysis or symbolic execution methods.

In summary, \textsc{StateGuard} proves its practicality in detecting defects in real-world smart contracts and demonstrates good adaptability to handle smart contracts of various sizes.

\begin{tcolorbox}[boxrule=1pt,boxsep=1pt,left=2pt,right=2pt,top=2pt,bottom=2pt]
\textbf{Answer to RQ3.}
\textsc{StateGuard} has proven effective in detecting state derailment defects in real-world smart contracts, as evidenced by its successful identification of several novel defects that received CVE IDs, which are undetected by other benchmark tools.
\end{tcolorbox} 

\section{Discussion} \label{sec:discussion}

In order to understand the state derailment defects in smart contracts, we perform a detailed statistical analysis of the selected DAppSCAN dataset in Table \ref{table:dataset}. As shown in Table \ref{tab:proportion}, we collect and count the percentage of each type of defect in the dataset. As indicated in Table \ref{tab:proportion}, based on the defect ratio analysis, logical inconsistency is the most important source of problems, accounting for 51.15\% of the state derailment defects, emphasizing the importance of logical design accuracy. Secondly, type and declaration errors and resource constraints are vital issues affecting contract execution efficiency and security. Exception handling and access control, although accounting for a lower percentage, are equally important in preventing potential security breaches. 
Therefore, the stability and security of smart contracts are crucial and depend on high-quality code. To ensure this, comprehensive strategies and measures must be implemented. 

\begin{table}[ht]
\centering
\caption{Defect Proportion.} 
\label{tab:proportion}
\begin{tabular}{@{}cc@{}}
\toprule
Cause of Defect & Proportion \\ \midrule
Logical Inconsistencies & 51.15\% \\
Type and Declaration Errors & 12.23\% \\
Resource Constraints & 20.55\% \\
Exception Handling & 6.84\% \\
Access Control & 9.23\% \\ \bottomrule
\end{tabular}
\end{table}

The practical implications of the discovered state derailment defects in DEXs are profound and multifaceted, impacting both the security and functionality of these platforms. Logical inconsistencies can lead to unintended contract behaviors, potentially resulting in financial losses for users due to erroneous state updates. Resource constraints highlight the necessity for efficient contract design to prevent gas limit overconsumption, which can disrupt contract execution and lead to denial-of-service scenarios. Access control defects expose DEXs to unauthorized state modifications, increasing the risk of fraud and asset theft. Type and declaration errors can cause state inconsistencies, undermining the reliability of the contract's operations and potentially leading to exploitations. Lastly, inadequate exception handling can result in unhandled errors, causing contract failures and opening avenues for denial-of-service attacks. Collectively, these defects underscore the critical need for rigorous security audits and robust contract design practices to ensure the integrity, reliability, and security of smart contracts within the DEX.

\subsection{Case Analysis} 
A DEX platform that utilizes smart contracts to enable an auto-liquidity mechanism~\cite{Xu_2023_SoK} for token transactions.
Fig. \ref{fig:code2} shows a simplified code snippet of the state derailment defect. We explain how an attacker exploits this defect and highlight the severe consequences that may cause state derailment.

In the ERC20 token standard, the \texttt{safeTransferFrom} function is a vital interface to transfer tokens between two addresses. The parameters of this function include the address of the token contract (\texttt{\_token}), the sender's address (\texttt{\_from}), the recipient's address (\texttt{\_to}), and the amount of tokens to be transferred (\texttt{\_value}). However, this function is defined as public, which means that anyone can call this function, potentially leading to some security issues.

\begin{figure}[htbp]
\centering
\includegraphics[width=\linewidth]{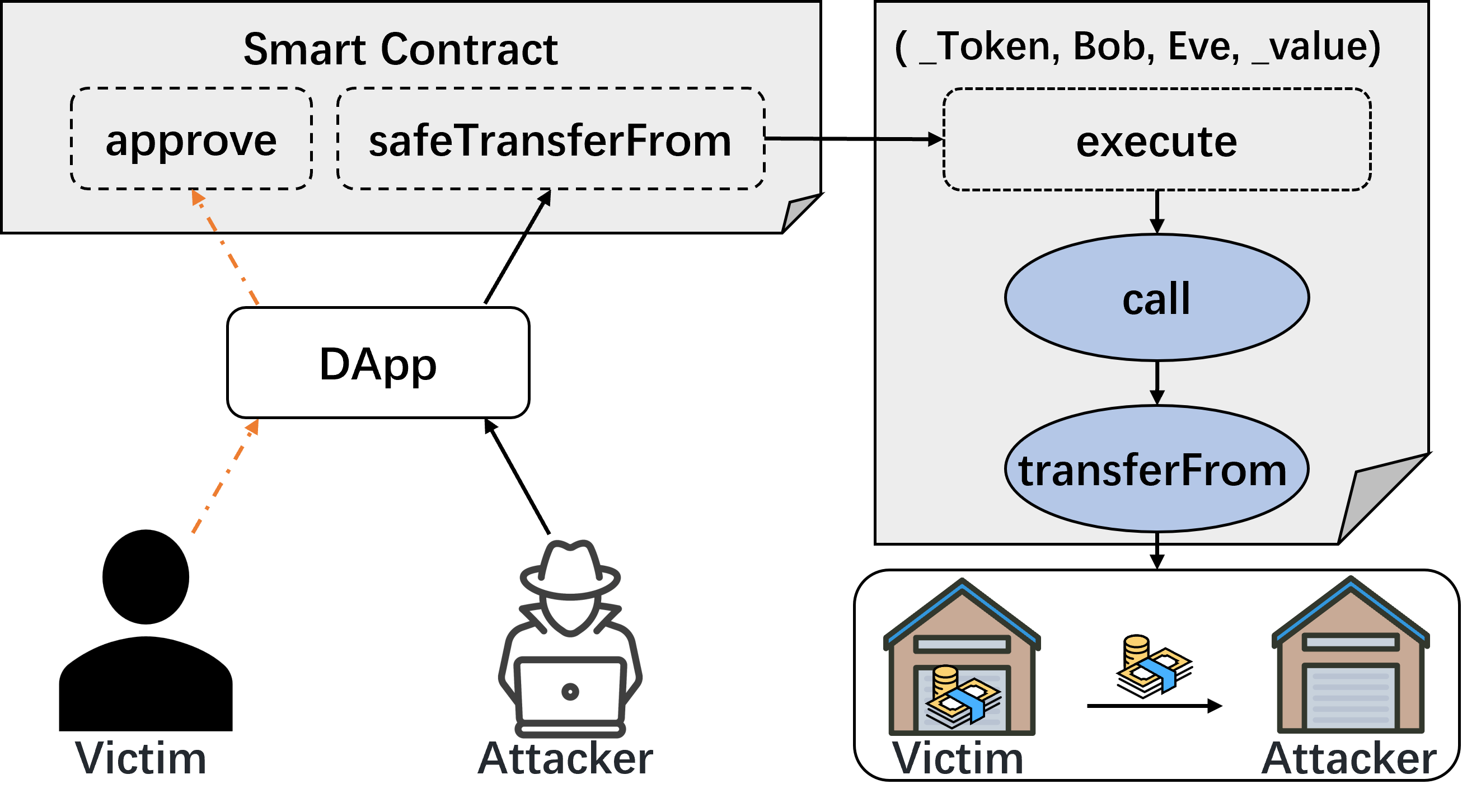}
\caption{Illustration of the State Derailment Case.}
\label{fig:bancor}
\end{figure}

To explain this issue in more detail, we can refer to an example in Fig. \ref{fig:bancor}. The \texttt{approve} function is another interface in the ERC20 standard, authorizing other addresses to transfer tokens on behalf of the \emph{Victim}. In this case, the \emph{Victim} calls the \texttt{approve} function and authorizes a certain amount of tokens for the platform's smart contract for trading on their platform. However, the platform's \texttt{safeTransferFrom} function does not restrict the identity of callers. Therefore, the \emph{Attacker} can exploit this defect by transferring tokens authorized by the \emph{Victim} to their address through the platform's smart contract without further consent from the \emph{Victim}.

Specifically, the \emph{Attacker} effectively operates proxy transfer. Proxy transfer is an operation in the blockchain, especially in the ERC20 token standard, where an authorized third party is allowed to transfer tokens from one account to another. Since there is no caller authentication within that particular function implementation, this allows successful execution where tokens will be transferred from the \emph{Victim} to the \emph{Attacker}.

\begin{figure}[htbp]
\setlength{\abovecaptionskip}{0.cm}
\begin{lstlisting}[numbers=none]
function safeTransferFrom(IERC20Token _token, address _from, address _to, uint256 _value) public {
   execute(_token, abi.encodeWithSelector(TRANSFER_FROM_FUNC_SELECTOR, _from, _to, _value));
}
\end{lstlisting}
\caption{Code Snippets of Defective Contracts.}
\label{fig:code2}
\end{figure}

\subsection{Limitations and Future Work}

Although we have made progress and optimizations in detecting DEX smart contracts, there remain two limitations. 

The time complexity of graphical representation is worth further optimization. DApps are structured with many smart contracts due to their project form. Therefore, the fan-out nature of AST results in a complex graphical representation, even though we have mitigated this problem using graph optimization methods. The time complexity for processing a subgraph is \(O(V + E)\), where \(V\) is the number of nodes and \(E\) is the number of edges. Suppose we must perform complex analysis tasks on the AST, such as data dependency or control dependency analysis. In that case, we may need to traverse the AST multiple times, increasing the time complexity.

Another area for improvement lies in data processing, which involves data extraction and selection. Specific account permissions and call relationships from contracts are critical in data extraction. However, it parses meaningful and highly correlated features from complex structured data. In data selection, noise and outliers widely exist in smart contracts, such as unused variables, dead code, and redundant code, which further increases the complexity of the task. Therefore, meticulous data processing is required to extract and define these features accurately.

Future work could incorporate more domain knowledge into graphical representation optimization and data processing, involving closer collaboration with field experts. Moreover, we plan to explore large language models (LLMs) for smart contract defect detection.

\section{Related Work} \label{sec:related work}

\subsection{Defect Detection Tools} 
 Researchers and developers have designed many defect detection tools in response to the potential defects in smart contracts. 
Solhint~\cite{Solhint} is an essential tool for ensuring code quality and consistency in smart contract compliance checking. 
 Static analysis tools~\cite{Yin_2022_empirical} (i.e., Mythril, Securify, Slither, and SmartCheck) offer robust defect detection
, complemented by dynamic analysis~\cite{Samreen_2021_SmartScan} from Manticore
Moreover, the comprehensive approach of MythX~\cite{MythX} combines static and dynamic analysis with fuzzing techniques. Deep learning and machine learning have also been utilized, with SaferSc~\cite{tann2019safer} using LSTM networks for defect detection, while Eth2Vec~\cite{ashizawa2021eth2vec} and ContractWard~\cite{wang2020contractward} leverage machine learning for defect detection in Ethereum Virtual Machine (EVM) bytecode.
Additionally, VulANalyzeR~\cite{VulANalyzeR} introduces a novel approach by combining sequential and topological learning through recurrent units and graph convolution, effectively simulating program execution to detect defects.  Meanwhile, TaintGuard~\cite{TaintGuard} stands out as a cross-contract static analysis tool designed to prevent implicit privilege leakage in Solidity smart contracts, utilizing taint analysis and instrumentation monitoring to filter call relations for cross-contract calls and detect problematic paths that may lead to privilege leaks.

\subsection{Security Analysis on DEX} 
A DEX permits users to trade encrypted assets directly via smart contracts, circumventing the need for traditional CEXs. The security of a DEX directly influences the safety of users' assets, thereby necessitating a thorough security analysis. A comprehensive security analysis typically encompasses auditing the DEX's smart contract, assessing the robustness of its design, and simulating attack scenarios~\cite{Zheng_2023_Blockchain-Based}. Since the DEX operates on the blockchain, attackers can exploit any security defect, potentially leading to significant financial losses. Therefore, the security of smart contracts in DEX is crucial for the safety of users' assets. For instance, Duan et al.~\cite{Duan_2022_Automated} proposed a program analysis technique, VetSC, capable of automatically extracting contract semantics from DApps and performing targeted security checks. VetSC can identify security risks in real-world DApps and ensure the security of decentralized applications. Conversely, Li et al.~\cite{Li_2021_SolSaviour} introduced SolSaviour, a defensive framework designed to repair and recover deployed flawed smart contracts. SolSaviour proposed a novel mechanism, termed the voteDestruct mechanism, which enables contract stakeholders to vote on the destruction of flawed smart contracts. In addition, Xia et al.~\cite{Xia_2021_Tradea} proposes a method based on the "Guilt-by-association" heuristic and machine learning techniques to identify fraudulent tokens on Uniswap from another perspective. This method can accurately label fraudulent behaviour on Uniswap. However, from a distinct perspective, Geoffrey et al.~\cite{Ramseyer_2023_SPEEDEXa} introduced  SPEEDEX to eliminate the prevalent front-running attacks in centralized exchanges and effectively parallelize transaction processing, achieving high throughput.

\section{CONCLUSION} \label{sec:conclusion}
In this paper, we present the first systematic study of state derailment defects in DEX smart contracts. We define and classify state derailment defects into five categories and provide examples and detailed analyses for each category. To discover security issues in DEX contracts, we design and develop \textsc{StateGuard}, a deep learning-based framework for detecting state derailment defects in DEX projects. We have evaluated \textsc{StateGuard} on two large datasets, i.e., DAppSCAN and Smartbugs. The results show that \textsc{StateGuard} identifies state derailment defects with 92.24\% precision and 90.4\% recall, outperforming several existing tools. Furthermore, \textsc{StateGuard} can discover defects in real-world contracts, demonstrating its practicality and effectiveness. As a next step, we plan to explore further leveraging LLM techniques to enhance our defect detection capabilities.

\section{ACKNOWLEDGMENTS}
This work is sponsored by the National Natural Science Foundation of China (No.62402146 \& 62362021).


 
%

\bibliographystyle{IEEEtran}
\bibliography{ref} 












\vfill

\end{document}